\documentclass[10pt,conference]{IEEEtran}

\usepackage{graphicx}
\usepackage{dcolumn}
\usepackage{array}
\usepackage{amsmath}
\usepackage{fancyheadings}
\usepackage{subfigure}
\usepackage{verbatim}
\usepackage{textcomp}
\usepackage{times,color}
\usepackage[final,ulem=normalem]{changes}

\newtheorem{definition}{Definition}
\newtheorem{observation}{Observation}

\definechangesauthor{RW}{blue}
\definechangesauthor{JW}{red}
\definechangesauthor{TYC}{magenta}
\definechangesauthor{TOM}{orange}
\definechangesauthor{BX}{cyan}

\begin{document}
\pagestyle{empty}

\title{Designing Nonlinear Turbo Codes \\ with a Target Ones Density}
\author{Jiadong Wang, Thomas Courtade, Tsung-Yi Chen, Bike Xie and Richard Wesel\\
wjd@ee.ucla.edu, tacourta@ee.ucla.edu, tychen@ee.ucla.edu, xbk@ee.ucla.edu, wesel@ee.ucla.edu}

\maketitle \thispagestyle{empty}

\begin{abstract}
Certain binary asymmetric channels, such as Z-channels in which one of the two crossover probabilities is zero,  demand optimal ones densities different from 50\%. Some broadcast channels, such as broadcast binary symmetric channels (BBSC) where each component channel is a binary symmetric channel, also require a non-uniform input distribution due to the superposition coding scheme, which is known to achieve the boundary of capacity region. This paper presents a systematic technique for designing nonlinear turbo codes that are able to support ones densities different from 50\%. To demonstrate the effectiveness of our design technique, we design and simulate nonlinear turbo codes for the Z-channel and the BBSC.  The best nonlinear turbo code is less than 0.02 bits from capacity.

\end{abstract}


\section{Introduction}\label{sec:intro}
Unlike their linear counterparts, nonlinear turbo codes are a family of turbo codes which can have average ones densities not equal to 50\%. The constituent code symbols of nonlinear turbo codes are not restricted to linear combinations of state and input bits, and any ones density can be achieved by using a look-up table that maps the state and input bits to output bits. Parallel concatenated nonlinear turbo codes have been designed to maintain specific ones densities for multiple access channels \cite{griot-GC06} and broadcast Z-channels \cite{Xie08}.

In \cite{griot-GC06,Xie08}, the look-up tables defining nonlinear turbo codes were designed in an ad-hoc manner, mostly by hand.  These techniques do not extend to trellises with many states.  This paper provides a new and efficient technique for designing nonlinear turbo codes.  The advantage of this technique is that the complexity does not grow exponentially with the number of states.  We demonstrate the effectiveness of the codes designed using our technique on both the point-to-point Z-channel and the broadcast binary symmetric channel (BBSC).

This paper is organized as follows.  Section \ref{sec:nlturbo} introduces nonlinear turbo codes and describes an efficient design scheme in a step-by-step manner. Section \ref{sec:results} describes nonlinear turbo codes designed specifically for the Z-channel and the BBSC and provides simulation results. Section \ref{sec:conclusions} delivers the conclusions.


\section{A Design Scheme for Nonlinear Turbo Codes}
\label{sec:nlturbo}
A nonlinear turbo code is defined by look-up tables that map state and input bits to output bits in the trellises of the constituent codes.  By controlling the number of ones in the look-up table, any desired ones density can be closely approximated.  In general, a brute-force search to find the look-up table yielding the largest effective free distance is impractical because the complexity grows exponentially with the number of states.

In this section, we propose a systematic method for designing nonlinear turbo codes with a complexity that does not grow exponentially in the number of states. Therefore, our approach can be used to design codes employing trellises with many states. In Section \ref{sec:results}, we use this technique to design several capacity-approaching nonlinear turbo codes for the Z-channel and the BBSC.  We would like to remark that although our technique is described in the context of binary turbo codes that use identical constituent encoders, it immediately extends to more general cases.

\subsection{State sub-tables, branch distance, and merge distance}\label{sec:definitions}
Throughout this section, we consider a trellis corresponding to a constituent code of the turbo code.  We assume that the trellis has $\ell$ states and each trellis transition corresponds to $k$ input bits and  $n$ output bits.  Using this notation, we define a state sub-table.

\begin{definition}
A {\bf state sub-table}  $M(s)$ corresponding to state $s$ is a $2^k$-by-$n$ binary matrix describing the mapping of input-bits to output-bits when the encoder is in state $s$ as follows.  When the encoder is in state $s$ and the input-bits are $b_1,\dots,b_k$, the encoder outputs the $n$ bits in the row corresponding to the $k$-tuple $b_1,\dots,b_k$. Without loss of generality, we assume that the rows of $M(s)$ are indexed by the binary $k$-tuples in lexicographical order.
\end{definition}

We also define the minimum branch-distance and the minimum merge-distance as follows:
\begin{definition}
The {\bf branch-distance} of state $s$ is the minimum distance between the rows of $M(s)$, the branch-distance of a trellis is the minimum of the state branch-distances.  The {\bf merge-distance} of a state and the merge-distance of a trellis are defined analogously as the minimum distances between the $n$-bit outputs of trellis transitions that merge into a state.
\end{definition}

Note that the particular distance metric can depend on the channel, as we will see in Section \ref{sec:results}.

As discussed above, designing the constituent codes for a nonlinear turbo code includes defining a look-up table that maps state and input bits to output bits in a trellis.  A look-up table that defines a constituent code with a large effective free distance is preferable to one that defines a constituent code with a small effective free distance.  A good heuristic for determining whether a look-up table will produce a code with a large effective free distance is to analyze the distances at the branches and merges of the trellis.  Our approach is based on the following key observation:

\begin{observation}
If $\Pi_1$ and $\Pi_2$ are $2^k \times 2^k$ and $n\times n$ permutation matrices respectively, then a state sub-table $M(s')=\Pi_1 M(s) \Pi_2$ has the same branch distance properties as $M(s)$ in the sense that the set of distances between outputs from state $s$ is the same as the set of distances between outputs from state $s'$.
\end{observation}

\subsection{Description of Design Scheme}\label{sec:designscheme}
With this observation in mind, we now describe our design approach:
\begin{enumerate}
\item If we require a constituent code with ones density $u_1$, choose $\nu$ to be the nearest integer to $u_1 \cdot n \cdot 2^k$.

\item Select parameters $d_b$ and $d_m$, where $d_b$ is the desired branch-distance of the trellis and $d_m$ is the desired merge-distance of the trellis.

\item Create a state sub-table $M(1)$ with exactly $\nu$ ones.  This ensures that the average ones density of the output bits from state $1$ is $\frac{\nu}{n \cdot 2^k}$ which is approximately $u_1$.  The state sub-table is designed by carefully placing the ones so that the branch-distance is greater than $d_b$.  If this is not possible, then return to step 2 and select a smaller $d_b$.  This step is the main source of complexity.  If $n$ and $k$ are sufficiently small, a good $M(1)$ can be found via a brute-force search.

\item For each other state $s\in \{2,\dots,\ell\}$, choose random permutation matrices $\Pi_1$ and $\Pi_2$ and set $M(s)=\Pi_1 M(1) \Pi_2$.  By Observation 1, this ensures that the trellis has branch-distance greater than $d_b$.

\item Check the resulting merge-distance of the trellis.  If it is less than $d_m$, return to step 4.  If a maximum number of iterations is reached, return to step 2 and select a smaller $d_m$.

\item Check the effective free distance of the obtained code.  Steps 3-5 are generally repeated several times to produce several candidate codes.  Usually, we select the code with the largest effective free distance.
\end{enumerate}

With this scheme, we can construct nonlinear turbo codes without an exhaustive computer search over all lookup tables with the desired ones density. Since we only manually design a sub-table for one state, the complexity does not grow exponentially in the number of states. Through a series of many experiments, we have observed that this procedure is effective for designing nonlinear turbo codes that approach capacity in channels demanding nonuniform ones densities.  In Section \ref{sec:results}, we use this technique to design nonlinear turbo codes for Z-channels and BBSCs.

\section{Examples}
\label{sec:results}
In this section, we apply our design technique to two different types of channels: Z-channels and BBSCs.  In each case, we successfully design nonlinear turbo codes that approach the capacity of their respective channels.  For the Z-channel, the codes are designed to perform well with respect to the directional Hamming distance metric, while in the case of the BBSC, the traditional Hamming distance metric is used.

\subsection{Z-channels: Introduction and directional distance}
\begin{figure}[t]
\centering
\includegraphics[width=0.25\textwidth]{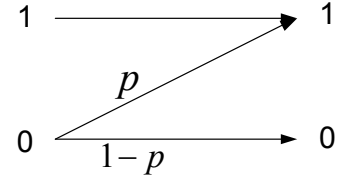}
\caption{Channel model for the Z-channel. }\label{fig:z_channel}
\vspace{0.5in}
\end{figure}

Z-channels are binary asymmetric channels in which one of the two crossover probabilities is zero (see Figure~\ref{fig:z_channel}).  This subsection considers point-to-point Z-channels and uses the model in which the probability of the $ 1 \rightarrow 0 $ crossover is 0. This channel applies to certain data storage systems \cite{Constantin1979} and certain optical communication systems \cite{McEliece1980}.

Golomb \cite{Golomb1980} studied the capacity and optimal ones density of Z-channels. The capacity of the Z-channel is given by
\begin{equation}
C=H(u_0(1-p))-u_0H(p)
\end{equation}
with optimal zeros density
\begin{equation}
u_0=\frac{p^{p/(1-p)}}{1+(1-p)p^{p/(1-p)}}.
\end{equation}

The optimal ones density for the Z-channel is higher than 50\% everywhere except for the noiseless channel where the crossover probability is zero.
All linear codes have an average ones density of 50\%, and are thus prevented from achieving the capacity of the Z-channel.  However, nonlinear turbo codes can provide any ones density and thus have the potential to achieve the capacity of the Z-channel.

For the Z-channel, directional Hamming distance, introduced in \cite{Constantin1979}, accurately describes the distance between codewords under Z-channel distortion. Consider two codewords of length-$n$ bits, $ X $ and $ \tilde{X} $. The directional Hamming distance between $ X $ and $ \tilde{X} $ is
\begin{equation}
d_D(X, \tilde{X} ) = \sum \limits_{1\leq i \leq n} I (x_i=0, \tilde{x}_i=1 )
\end{equation}
where $ I $ is the indicator function. The directional Hamming distance is asymmetric, and $ d_D(X, \tilde{X} ) $ is usually not equal to $ d_D(\tilde{X}, X) $.  A more precise definition of the pair-wise distance of Z-channel is defined in \cite{griot-GC06} and \cite{Klove}:
\begin{equation}
d_Z(X, \tilde{X} ) = \text{max}\left[ d_D(X, \tilde{X} ), d_D(\tilde{X}, X) \right]
\end{equation}
since the larger directional distance matters in the decoding.

\subsection{Z-channels: Code Design}
The design begins with the 16-state duo-binary trellis described in \cite{Xie08}. With the design technique described in Subsection~\ref{sec:designscheme}, we first design a systematic rate-1/10 turbo code with a {\em target} ones density of 0.621.  The actual ones density produced is 0.5953. This rate-1/10 code maximizes the minimum pairwise directional Hamming distance for the splits from each state and the merges to each state.

We choose the best code with largest directional effective free distance among the randomly generated nonlinear turbo codes. Table~\ref{tab:NLTC-label} shows the nine nonsystematic (parity) bits for the two identical constituent nonlinear binary trellis encoders that comprise this turbo code.  The two constituent encoders each produce nine parity bits, which join with the two systematic bits to produce 20 coded bits for each two-bit input symbol.  The interleaver for this turbo code can be found in \cite{link}.

As is commonly done with turbo codes, we puncture this rate-1/10 code to create a variety of rates. The puncturing patterns are given in Table \ref{tab:Puncture-low}  with the code rates and ones densities they produce.  In each case, the resulting ones density closely approximates the optimal ones density.

\begin{table}
\begin{center}
\caption{Octal labeling for constituent binary trellis codes. Rows represent the state $s_1 s_2 s_3 s_4$, columns represent the input $u_1 u_2$. }\label{tab:NLTC-label}
\begin{tabular}{|c||c|c|c|c|}
\cline{1-5} state & \multicolumn{4}{c|}{input}\\
\cline{2-5}
& 00 & 01 & 10 & 11 \\
\cline{1-5}
0000 & 534	& 343	& 671	& 517	\\
\cline{1-5}
0001 & 476	& 073	& 707	& 364	\\
\cline{1-5}
0010 & 346	& 257	& 571	& 632	\\
\cline{1-5}
0011 & 137	& 752	& 711	& 265	\\
\cline{1-5}
0100 & 754	& 566	& 227	& 171	\\
\cline{1-5}
0101 & 370	& 467	& 516	& 335	\\
\cline{1-5}
0110 & 743	& 574	& 037	& 626	\\
\cline{1-5}
0111 & 566	& 273	& 532	& 615	\\
\cline{1-5}
1000 & 465	& 457	& 343	& 334	\\
\cline{1-5}
1001 & 752	& 665	& 037	& 370	\\
\cline{1-5}
1010 & 274	& 563	& 754	& 307	\\
\cline{1-5}
1011 & 723	& 354	& 617	& 465	\\
\cline{1-5}
1100 & 435	& 643	& 317	& 564	\\
\cline{1-5}
1101 & 153	& 666	& 703	& 334	\\
\cline{1-5}
1110 & 327	& 176	& 453	& 664	\\
\cline{1-5}
1111 & 466	& 153	& 335	& 761	\\
\cline{1-5}
\end{tabular}
\end{center}
\end{table}


\begin{table}
\begin{center}
\caption{Octal puncturing patterns (in octal) of parity bits. Punctured bits are indicated by 1's. The puncturing period is 9 bits.}\label{tab:Puncture-low}
\begin{tabular}{|c||c|c|c|c|}
\cline{1-5} Rate & Encoder 1 & Encoder 2 & Ones Density & Optimal Density \\
\cline{1-5}
1/10 & 000	& 000	& 0.5953 & 0.621	\\
\cline{1-5}
1/9 & 001	& 002	& 0.5955 & 0.6197 \\
\cline{1-5}
1/8 & 201	& 042	& 0.5938 & 0.6181	\\
\cline{1-5}
1/7 & 241	& 043	& 0.5915 & 0.6161	\\
\cline{1-5}
1/6 & 243	& 243	& 0.5911 & 0.6134	\\
\cline{1-5}
1/5 & 247	& 263	& 0.5828 & 0.6094	\\
\cline{1-5}
1/4 & 257	& 267	& 0.5742 & 0.6035	\\
\cline{1-5}
1/3 & 277	& 367	& 0.5599 & 0.5931	\\
\cline{1-5}
\end{tabular}
\end{center}
\end{table}


\subsection{Z-channels:  Numerical Results}
\begin{figure}
\centering
\includegraphics[width=0.4\textwidth]{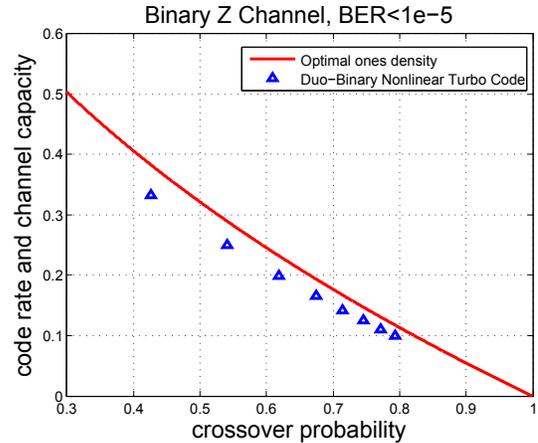}
\caption{The capacity with optimal ones density and designed nonlinear Turbo codes for Z-channels.  The operating points from left to right correspond to the rate-1/3 to rate-1/10 nonlinear turbo codes respectively.}
\label{fig:z_rate}
\end{figure}

We simulated our nonlinear turbo codes on Z-channels with capacities slightly greater than the code rates. All of the simulations use 20,000 input bits per codeword and an extended spread interleaver \cite{FragouliTC01}.
The complete interleaver description is available online \cite{link}. Figure~\ref{fig:z_rate} shows the capacity of the Z-channel for different crossover probabilities and the observed operating point
(code rate and crossover probability) where each codes achieved bit error rates of less than $10^{-5}$.  The distance from capacity ranges from approximately $0.018$ to $0.05$ bits.

\subsection{BBSC: Introduction}
\begin{figure}[t]
\centering
\includegraphics[width=0.4\textwidth]{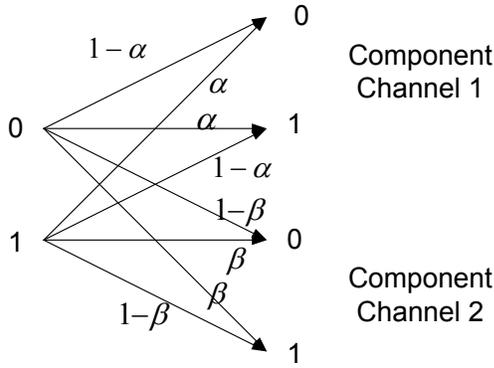}
\caption{Channel model for two-user broadcast binary-symmetric channel}\label{fig:bbsc_channel}
\vspace{1in}
\end{figure}

\begin{figure}[t]
\begin{center}
\includegraphics[width=0.4\textwidth]{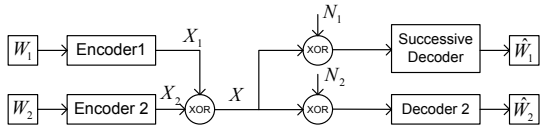}
\caption{Encoder scheme for two-user broadcast binary-symmetric channel}
\label{fig:bbsc_encoder}\end{center}
\end{figure}

We also designed nonlinear codes for the two-user BBSC, which consists of two binary symmetric component channels, one with transition probability $ \alpha $ and the other with transition probability $ \beta $, as shown in Figure~\ref{fig:bbsc_channel}. Without loss of generality, we assume $ \alpha < \beta $. A simple and optimal encoding scheme is an independent-encoding approach in which symbols from independent codebooks are added together using the XOR function. We refer to this scheme as superposition coding, and the encoder structure is shown in Figure~\ref{fig:bbsc_encoder}.

The capacity region of a degraded broadcast channel was established by Cover \cite{Cover1972}, Bergmans \cite{Bergmans1973} and  Gallager \cite{Gallager1974}. Cover \cite{Cover1975} introduced an independent-encoding scheme for two-user broadcast channels. This scheme is known to achieve the boundary of the capacity region for the broadcast binary-symmetric channel (BBSC) and is investigated in \cite{Wyner1973} \cite{Witsenhausen1974} \cite{Kasami1985} \cite{BhatGlobecom10}.

The capacity region for a BBSC is given by
\begin{equation}\label{equ_capBBSC}
\begin{split}
R_1  & \le  h ( \alpha \ast p_1) - h ( \alpha ) \\
R_2  & \le  1 - h ( \beta \ast p_1),
\end{split}
\end{equation}
where $ p_1 $ is the ones density of $ X_1 $, $ X_2 $ has 50\% ones density and the operation $ \ast $ is defined by
\begin{equation}
a \ast b = a(1-b)+b(1-a), \quad 0<a,b<1.
\end{equation}

In order to use a superposition coding scheme, the codes of the two users cannot both have ones densities of 50\%.  This precludes the exclusive use of linear codes.
Using the techniques described in Section \ref{sec:designscheme}, we design a family of nonlinear turbo codes that can provide a controlled ones density.
Superposition of one of our nonlinear codes with a linear turbo code produces an overall transmission with the potential to approach an optimal point the capacity region of the BBSC.

\subsection{BBSC: Code Design}
This section describes the construction of practical superposition codes for the two-user BBSC under different channel scenarios.
For the second user, we use a linear turbo code from the DVB-RCT standard, which has a 16-state duo-binary turbo encoder trellis structure, and extend it to lower rates.
For the first user we use nonlinear turbo codes with various ones densities, a 16-state duo-binary turbo encoder trellis structure from \cite{Berrou2005} when the channel parameters of  Fig. \ref{fig:bbsc_channel} are
$(\alpha,\beta)=(0.188,0.2017)$, and the 64-state 4-input turbo encoder trellis structure shown in Figure~\ref{fig:encoder64} when $(\alpha,\beta)=(0.01,0.108)$. The design process is described as follows:

\begin{figure}
\centering
\includegraphics[width=0.4\textwidth]{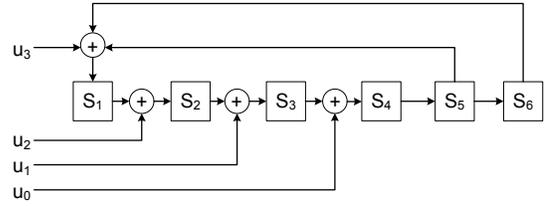}
\caption{64-state 4-input turbo encoder trellis structure}
\label{fig:encoder64}
\vspace{0.5in}
\end{figure}

\begin{enumerate}
\item Choose an appropriate ones density for the first user. As shown in \eqref{equ_capBBSC}, for a BBSC with fixed $ \alpha $ and $ \beta $, the boundary of the capacity region is a function of $ p_1$, the ones density of User 1:
\[ R_1  =  f_1(p_1),  R_2 =  f_2(p_1) .\]
where $ f_1 $ and $ f_2 $ are monotonically increasing and decreasing functions of $ p_1$ respectively. Since these functions are one-to-one for $ 0<p_1<0.5 $, $ p_1 $ can be calculated by
\[ p_1 = f_1^{-1}(R_1) = f_2^{-1}(R_2). \]

Thus under this channel, for any working point $ (r_1,r_2) $, where $ r_1 $ and $ r_2 $ are the rates of user 1 and user 2, pick the ones density $ p_1 $ that satisfies
\[ f_2^{-1}(r_2) < p_1 < f_1^{-1}(r_1) \]
where $ f_1^{-1}(r_1) $ ( $ f_2^{-1}(r_2) $ ) corresponds to the ones density of the intersection between the horizontal (vertical) line from $ (r_1,r_2) $ and the capacity curve.

\item For the desired $p_1$, use the approach in Section \ref{sec:designscheme} to design the look-up table for the nonlinear turbo code. The look-up tables, interleavers and puncture patterns could be found online \cite{link}.

\item If a low ones density and a high rate is needed, there will be multiple all zero outputs in the look-up table, which is problematic. Thus we design a lower rate codes with the same ones density by using step 2 above. Then we uniformly puncture the codewords to obtain the desired rate while preserving the desired rate.
\end{enumerate}

\begin{figure}[t]
\centering
\includegraphics[width=0.4\textwidth]{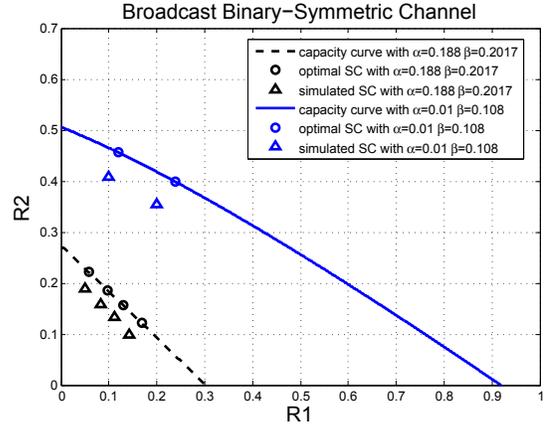}
\caption{Achievable capacity region and simulated points for superposition codes and time-sharing schemes for BBSC}
\label{fig:bbsc_simu}
\end{figure}


\begin{figure}[t]
\centering
\includegraphics[width=0.4\textwidth]{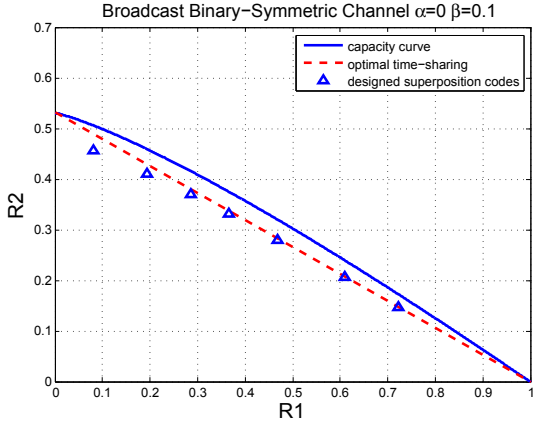}
\caption{A special case $ \alpha=0, \beta=0.1 $: capacity region and simulated points for superposition codes and time-sharing schemes for the BBSC}
\label{fig:bbsc_spec}
\vspace{0.5in}
\end{figure}

\subsection{BBSC: Numerical Results}
We simulated the superposition codes employing our nonlinear turbo codes with different rates for a variety of BBSCs. Figure~\ref{fig:bbsc_simu} shows the capacity region for 2 different channel scenarios and four operating points for each case (each operating point has a BER less than $ \leq 10^{-5} $).

All of the simulations use 20,000 input bits and an extended spread interleaver. For the nonlinear codes,  we use an $N=10000$ symbol-wise interleaver in the case where $(\alpha,\beta)=(0.188,0.2017)$, and we use an $N=5000$ symbol-wise interleaver in the case where $(\alpha,\beta)=(0.01,0.108)$. For the linear code, we use an $N=10000$ symbol-wise interleaver.

One special case of the BBSC is when $(\alpha,\beta)=(0,0.1) $.  This corresponds to the case when the first channel is noiseless and the distortion observed by User 1 is solely due to interference from User 2's superposed codeword. The capacity region in this scenario is
\begin{equation}
\begin{split}
R_1  &\le h(p_1 ) \\
 R_2  &\le 1 - h(\beta * p_1 ).
\end{split}
\end{equation}

Since the first channel is noiseless, we can use permutation codes \cite{Datta1999} whose rate can achieve $ h(p_1) $ for given ones density $ p_1 $, and traditional turbo codes for the second user. Simulation results are given in Figure~\ref{fig:bbsc_spec}.  We would like to point out that some of the operating points perform at rates higher than those theoretically achievable by time-sharing schemes.



\subsection{Remarks}
Linear turbo codes with 50\% ones density can also work under asymmetric channels with a loss in performance. In some channels, this loss could be small due to the small gap between the capacity with optimal ones density and mutual information with 50\% onesd density. However, for certain channels such as the multiple access OR channel, the theoretical gap is large and this new design technique of efficiently constructing nonlinear turbo codes is useful especially for a large trellis.

\section{Conclusions}
\label{sec:conclusions}
This paper proposes a systematic method for designing nonlinear turbo codes with a desired ones density. Using our approach, 
we designed a series of nonlinear turbo codes that approached  capacity in several Z-channels.  These codes were designed 
using the directional Hamming distance metric.  We also designed a series of nonlinear turbo codes which were scombined via superposition with 
linear turbo codes for use on the two-user BBSC.  These codes were designed using the traditional Hamming distance metric and 
also performed near capacity. The presented design techniques for nonlinear turbo codes could also be extended to general 
group-addition degraded broadcast channels \cite{Xie-ISIT09} as future work.



\bibliographystyle{ieeetran}
\bibliography{nltc}

\end{document}